\documentclass[preprint,aps,pra,showpacs]{revtex4}
\usepackage{tabularx}
\usepackage{dcolumn}
\usepackage{graphics}
\usepackage{bm}
\usepackage[dvips]{graphicx}
\usepackage[dvips]{rotating}
\usepackage[latin1]{inputenc}
\usepackage{dcolumn}
\usepackage{tabularx}
\usepackage{amsmath}
\usepackage{amssymb}
\begin{document}
\draft

\title{Modifying the photodetachment near a metal surface by a weak electric field}

\author{B. C. Yang and  M. L. Du}
\email{duml@itp.ac.cn} \affiliation{Institute of Theoretical
Physics, Academia Sinica, Beijing 100190, China}



\date{\today}

\begin{abstract}

We show the photodetachment cross sections of H$^-$ near a metal surface can be modified using a weak static electric field. The modification is possible because the oscillatory part of the cross section near a metal surface is directly connected with the transit-time and the action of the detached-electron closed-orbit which can be changed systematically by varying the static electric field strength. Photodetachment cross sections for various photon energies and electric field values are calculated and displayed.

\par


\end{abstract}

\maketitle

\section{Introduction}

Great interest in the photodetachment of negative ion in electric and magnetic fields has grown since 1980s when scientists observed oscillatory photodetachment cross sections for
negative ions in a static electric field \cite{Fabrikant1, Bryant1,
Stewart,Rau1, Du1, Gibson1, Peters,
Blondel, Rangan, Gibson2, Fabrikant2, Rau2, Kramer, Du2,
Bracher, Du3}. In the same time, energy level shifts and other interesting dynamics for Rydberg atom near a metal surface
have also been studied \cite{Ganesan, Simonovic1, Salas, Simonovic2, Wang}. Recently Yang \emph{et al} studied the photodetachment process of H$^-$ near an interface\cite{Yang}. Inspired by these studies, the photodetachment process of a negative ion near a metal surface has been proposed and studied recently by Zhao and Du\cite{Zhao1,Zhao2}. They showed that the image charge inside the metal induces an oscillation in the photodetachment cross section above threshold. Further more, it was predicated that the oscillation exists only when the detached-electron energy is less than $\frac{1}{4d}$, where $d$ is the distance between the initial negative ion and the metal surface. Below photodetachment threshold the quantum tunneling effect makes the photodetachment cross section finite.

In this article, we consider the coherent control of the photodetachment process of a negative ion such as H$^-$ near a metal surface by using an additional static electric field. We will show by adding a static electric field the photodetachment cross section will become oscillatory above threshold and the energy limit mentioned above for the oscillation in the presence of only a metal surface will be removed. In addition, when the static electric field is applied, the photodetachment cross section is controlled by the transit-time and the action of one detached-electron closed-orbit. Because the static electric field can systematically change the transit-time and the action of the closed-orbit, it can modify the photodetachment cross section in a systematical way. Atomic units will be used unless specified otherwise.

\section{Hamiltonian and closed-orbit}

The system is shown schematically in Fig.1. A negative ion H$^-$ is near a metal surface. An external static electric field $\textbf{F}$ perpendicular to the metal surface is applied. We assume the z-axis is away from the metal surface and the photon
polarization direction is also in the z direction. Using the image
method\cite{Ganesan,Zhao2}, the Hamiltonian for the detached-electron in cylindrical coordinates can be written as
\begin{equation}\label{}
     H=\frac{1}{2}(p_\rho^2+p_z^2)-\frac{1}{4(d+z)}+\frac{1}{4d}+Fz.
\end{equation}
The first four terms in Eq.(1) describe the motion of the detached-electron near a metal surface\cite{Zhao1,Zhao2}. The last term is new and it represents the static electric field. Note the constant term $\frac{1}{4d}$ does not change the dynamics. It is added to set the value of the total potential zero at the origin.

According to the closed-orbit theory\cite{Du4,Du5}, the photodetachment process can be described as follows.
The negative ion H$^-$ is initially in an $s$ state and the active electron is loosely bound by the hydrogen atom. When
the laser is on, the negative ion may absorb a photon of energy $E_{ph}$ and the active electron becomes an outgoing p-wave. The wave then propagates away from the hydrogen atom in all directions. When the distance between the
detached electron and the hydrogen atom is large,
a semiclassical description for the electronic motion is
appropriate. If there are closed-orbits in the system, the detached-electron wave can follow the closed-orbits and returns to the initial negative ion region. When this happens, the returning waves will interfere with the initial outgoing wave to produce oscillations in the photodetachment cross section. Following earlier study, it is assumed that the metal absorbs the electron when the electron hits the metal\cite{Zhao1,Zhao2}. In Fig.2 the photodetachment process and the potential for the Hamiltonian in Eq.(1) are illustrated.
Previously we have demonstrated that very close to photodetachment threshold the metal surface acts like an effective electric field of strength $F_{eff}$ defined by\cite{Zhao2}
\begin{equation}\label{}
    F_{eff}=\frac{1}{4d^2},
\end{equation}
where d is the distance between the metal surface and the negative ion.
In this article we will first fix $d=60a_0$ and discuss the modifications in the cross sections by varying the static electric field. The effect of changing the distance will be briefly discussed later. In Fig.2 the potentials in Eq.(1) are displayed for static electric field $F=0F_{eff}, 0.3F_{eff},
0.6F_{eff}$ and $0.9F_{eff}$ respectively.
The heavy black curve labeled by $F=0F_{eff}$ is the image potential of the metal surface without the electric field considered previously\cite{Zhao2}.  In this case, we found that if the detached electron energy is less than $\frac{1}{4d}$, there is a closed-orbit in the system. If the detached electron energy is greater than $\frac{1}{4d}$, there is no closed-orbit. According to closed-orbit theory, the photodetachment cross section is oscillatory when the detached-electron is above threshold and less than $\frac{1}{4d}$. The oscillatory cross section merges to a smooth cross section
for detached-electron energy equal to $\frac{1}{4d}$.
When the static electric field is present, the potential keeps increasing as z is increased. We find there is always one closed-orbit above threshold. As illustrated in Fig.2, this closed-orbit leaves the negative ion in the z-direction and finally returns to the negative ion after it is turned around by the image potential of the metal surface and the static electric field. Fig.2 also suggests the transit time of the closed-orbit decreases as the static electric field $F$ is increased. Numerical calculations will confirm this observation later.

\section{formulas for photodetachment cross section}

When the static electric field is on, the same procedure\cite{Zhao2} can be used to derive the photodetachment cross section. We will briefly summarize the results. For detached-electron energy $E\geq 0$, the photodetachment cross section is a sum of two terms,
\begin{equation}\label{}
    \sigma(E,d,F)=\sigma_0(E)+\sigma_{r}(E,d,F),~~E\geq0
\end{equation}
where
\begin{equation}\label{}
    \sigma_0(E)=\frac{16\pi^2\sqrt{2}B^2E^{3/2}}{3c(E_b+E)^3}
\end{equation}
is the smooth background and is equal to the cross section of free negative ion without the metal surface and the static electric field, and  $\sigma_r(E,d,F)$ is the oscillating part of the cross section given by
\begin{equation}\label{}
    \sigma_r(E,d,F)=\frac{8\pi^2B^2\sqrt{2E}}{c(E_b+E)^3T(E,d,F)}\cos[S(E,d,F)]
\end{equation}
where $T(E,d,F)$ and $S(E,d,F)$ are respectively the transit time and the action of the closed-orbit. The transit time and the action can be calculated using the following integrals
\begin{eqnarray}\label{}
    T(E,d,F)&=& 2\int_0^{z_{m}}\frac{1}{p_z}dz,\nonumber\\
    S(E,d,F)&=& 2\int_0^{z_{m}}p_zdz.
\end{eqnarray}
The momentum $p_z$ in the z-direction is readily obtained from Eq.(1) as
\begin{equation}\label{}
     p_z=\sqrt{2(E-\frac{1}{4d}+\frac{1}{4(d+z)}-Fz)}.
\end{equation}
$z_{m}$ is the turning point of the closed-orbit and can be obtained by setting $p_z$ to zero. If we denote the parameter $A=E-\frac{1}{4d}$, then
\begin{equation}\label{}
    z_m=\frac{1}{2F}[-(Fd-A)+\sqrt{(Fd+A)^2+F}].
\end{equation}
In both Eq.(4) and Eq.(5) $B=0.31552$, $E_b$ is the binding energy of H$^-$ and is approximately $0.754$eV, c is approximately equal to 137 a.u. \cite{Du1}. Both integrals for $T(E,d,F)$ and $S(E,d,F)$ can be expressed analytically using special functions as\cite{table}
\begin{equation}\label{}
    T=\frac{2\sqrt{2}}{\sqrt{F}}[\sqrt{z_m-z_n}E(\gamma,\lambda)+\frac{d+z_n}{\sqrt{z_m-z_n}}F(\gamma,\lambda)]
\end{equation}
and
\begin{equation}\label{}
    S=\frac{4\sqrt{2F}}{3}\{\sqrt{z_m-z_n}[(z_m+z_n+2d)E(\gamma,\lambda)-(z_n+d)F(\gamma,\lambda)]-\sqrt{-z_mz_nd}\}~,
\end{equation}
where $z_n<0$ is another point corresponding to $p_z=0$ in Eq.(7) and is given by
\begin{equation}\label{}
    z_n=\frac{1}{2F}[-(Fd-A)-\sqrt{(Fd+A)^2+F}]~.
\end{equation}
$F(\gamma,\lambda)$ and $E(\gamma,\lambda)$ (This function should not be confused with the energy $E$.) are, respectively, the
elliptic integrals of the first kind and the second kind\cite{table}. The two parameters are defined as
\begin{eqnarray}\label{}
    \gamma &=& \arcsin\sqrt{\frac{z_m}{z_m+d}}~,0\leq\gamma\leq\frac{\pi}{2};\nonumber\\
    \lambda &=& \sqrt{\frac{z_m+d}{z_m-z_n}}~,0\leq\lambda\leq 1.
\end{eqnarray}
Although the analytical expressions for $T(E,d,F)$ and $S(E,d,F)$ are derived above, in practice, we find it is easier to evaluate the  integrals in Eqs.(6) numerically.
For detached electron energy $E<0$, the cross section is finite because of a quantum tunneling effect. Following the earlier work\cite{Zhao2}, the above threshold photodetachment cross section joins to the following formula below threshold
\begin{equation}\label{}
    \sigma(E,d,F)=\frac{\pi^2B^2(1+4Fd^2)}{cd^2(E_b+E)^3}\exp[-2S_t(E,d,F)],~~E\leq0,
\end{equation}
where
\begin{equation}\label{}
    S_t(E,d,F)=\int^0_{z_{t}}\sqrt{2(Fz-\frac{1}{4(d+z)}+\frac{1}{4d}-E)}dz,~~E\leq0
\end{equation}
and the formula for $z_t$ has the same expression as $z_m$ in Eq.(8)
except the energy $E$ is negative.

\section{modifying cross sections}

We have expressed the cross section in terms of the transit time and the action of the only closed-orbit for the system in Eq.(1). We now fix the distance $d$ between the negative ion and the metal surface to be $60a_0$ and study how the static electric field modifies the transit time and the action of the closed-orbit. In Fig.3 we show the dependence of $T(E,d,F)$ and $S(E,d,F)$ on the detached electron energy for static electric fields $F=0F_{eff},0.3F_{eff}, 0.6F_{eff}$ and $0.9F_{eff}$, respectively. We calculated the transit time $T(E,d,F)$ and the action $S(E,d,F)$ independently. The numerical results were checked against the following relationship\cite{Du5}
\begin{equation}\label{}
    \frac{\partial S}{\partial E}=T.
\end{equation}

As the electric field strength is increased, both the transit time and the action of the closed orbit decrease. The changes in the transit time and the action of the closed orbit are directly reflected in the total photodetachment cross section. Simple analysis of the formula
in Eq.(5) suggests the oscillation amplitude is increased but the  oscillation frequency is decreased as the static electric field strength $F$ in Eq.(1) is increased. In Fig.4 we display the calculated cross sections corresponding to $F=0F_{eff},0.3F_{eff}, 0.6F_{eff}$ and $0.9F_{eff}$. The results are consistent with our expectations. In the presence of the static electric field the photodetachment cross section  is oscillatory above threshold. The oscillation amplitude becomes larger while the oscillation frequency becomes smaller as the electric field is increased.

Regarding the variation of the action $S$ with respect to the static electric field in the present system, we can apply a theorem\cite{Du5} to get
\begin{equation}\label{}
    \frac{\partial S}{\partial F}=-\int z dt.
\end{equation}
The integral on the right side of Eq.(16) is along the closed-orbit. The first derivative $\frac{\partial S}{\partial F}$ is therefore negative because the right side is negative for this closed-orbit.
As a result, when the static electric field $F$ is increased, the action of the closed-orbit always decreases. This dependence of action on the static electric field can be used to modify the cross section. For selected photon energies, we show in Fig.5 the variations of the cross sections as the static electric field is increased. The selected photon energies correspond to the minima in the cross section when the static electric field is zero. The photon energies for the dotted curve, the dashed curve and the heavy black curve are respectively $0.8533$eV, $0.8564$eV and $0.8585$eV, for example. Fig.5 demonstrates that for any fixed photon energy, the cross section varies in a simple pattern as the static electric field is increased. Therefore we can get the desired cross section by choosing  proper values of static electric field. For example, when the photon energy is $0.8585$eV, the cross section reaches a large value if the static electric field is close to 5kV/cm and the cross section reaches a small value if the static electric field is close to 3kV/cm.
In fact, we find the valleys and peaks of the oscillation in the cross section are primarily determined by the action $S$. To demonstrate this point, in Fig.6 we show $\cos(S)$ as a function of both energy and electric field. The brightest points correspond to the maxima of $\cos(S)$. For the purpose of comparison, we find the cross section represented by the black line in Fig.5 is similar to the brightness along the dotted line in Fig.6. The three peaks indicated by the circles in the cross section in Fig.5 coincide well with the three brightest spots indicated by the three circles in Fig.6. Because of this correspondence, Fig.6 can be used as a map for the modification of the photodetachment cross section.

Finally we emphasize that although the above analysis is made for a distance $d=60a_0$ between the negative ion and the metal surface, similar phenomena occur at other distances. Fig.7 shows the photodetachment cross sections for three distances with the same static electric field at 107kV/cm. Similar analysis for the interference patterns in the cross sections can be carried out for each distance to find the positions of the minima and maxima in the cross sections.

\section{Conclusions}

We have investigated the effects of an additional weak static electric field on the photodetachment of H$^-$ near a metal surface. When the static electric field is applied, the photodetachment cross section becomes oscillatory in the whole energy region above threshold. The cross section can be expressed in terms of the transit time and the action of the only closed-orbit. By increasing the static electric field, one can systematically change the transit time and the action of the closed-orbit to modify the cross section. The landscape of the function $\cos(S)$ can serve as a map for the modification of the photodetachment cross section.

\newpage

\begin{center}

\begin{figure}
\includegraphics[width=450pt]{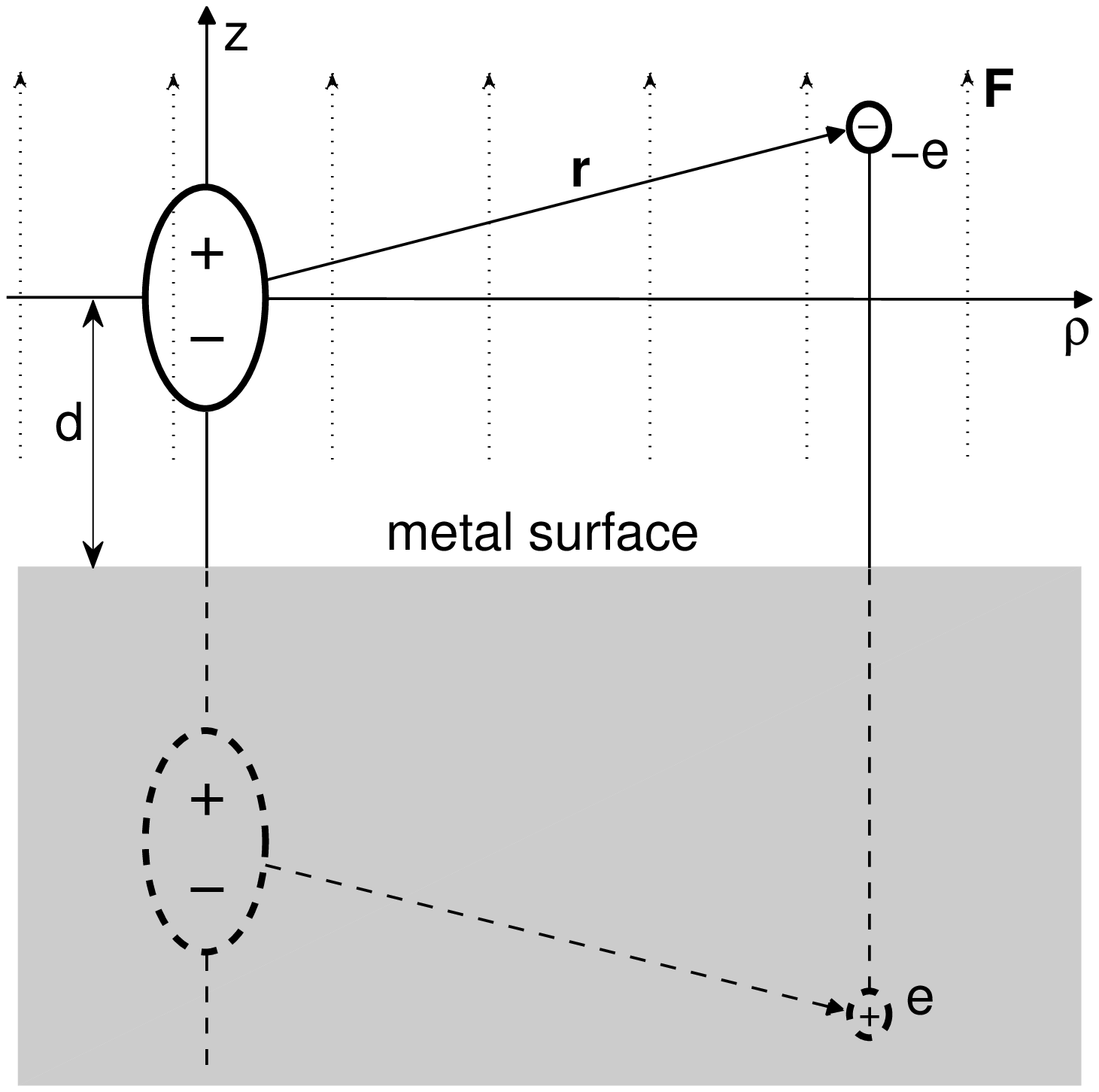}
\caption{Schematic representation of a hydrogen atom and the detached electron in the presence of a metal surface and a static electric field. The distance between the negative ion and the metal surface is d and the electric field points away from the surface. The detached electron moves in the image potential of the metal and the electric field. }
\end{figure}

\begin{figure}
  \includegraphics[width=450pt]{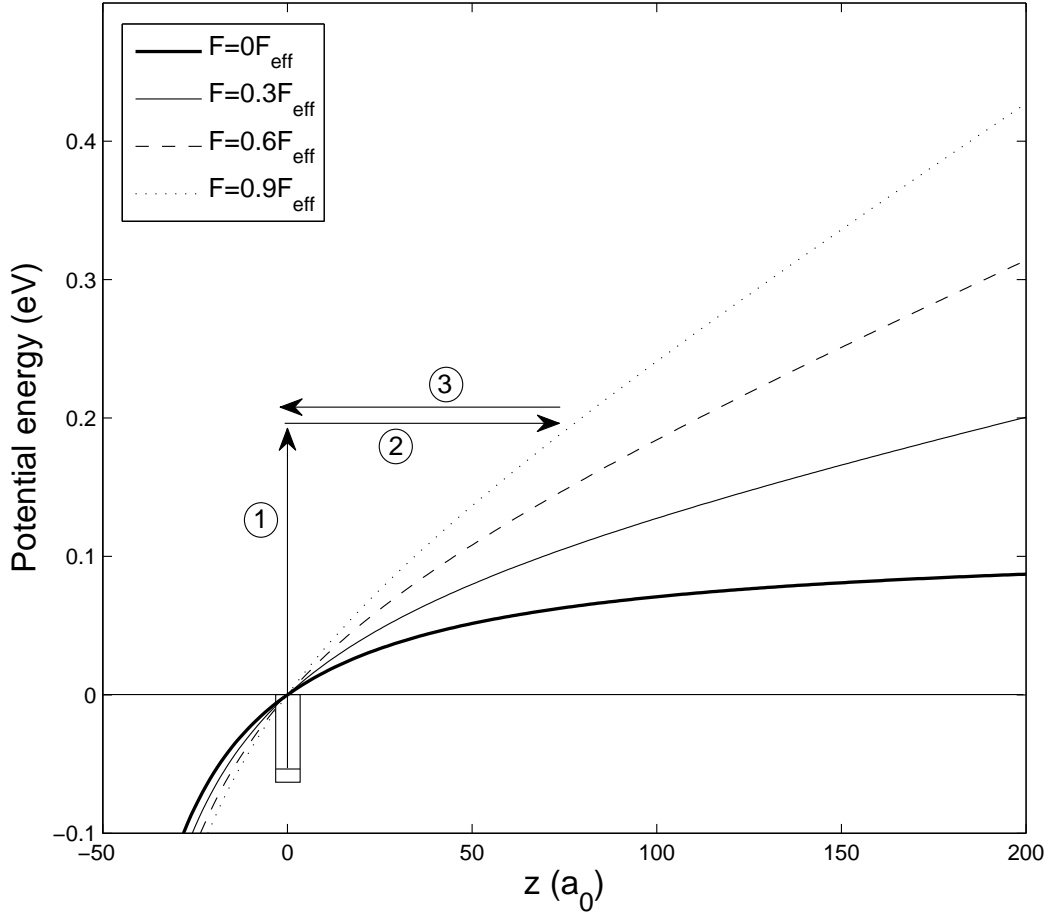}
  \caption{The potentials in Eq.(1) for several electric fields. The distance $d$ between the negative ion and the metal surface is $60a_0$.
  The effective field by the metal surface is defined by $F_{eff}=\frac{1}{4d^2}$. The photodetachment process can be described as follows: in the first step, the negative ion absorbs a photon and an outgoing detached electron wave is created; in the second step, the detached electron wave propagates away from the metal surface; in the third step, part of the detached electron wave
  is turned around and returns to the negative ion, where it interferes with the outgoing wave to produce oscillations in the cross section.  }
\end{figure}

\begin{figure}
  \includegraphics[width=450pt]{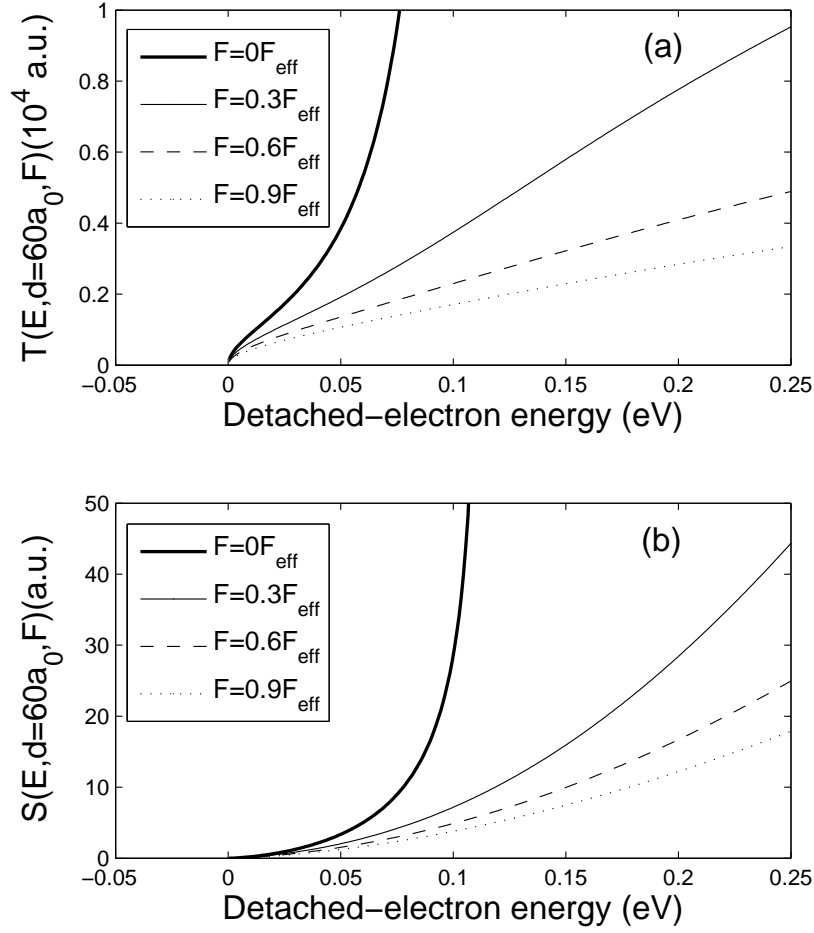}
  \caption{The transit time $T$ and the action $S$ of the closed-orbit.   The distance between the negative ion and the metal surface is $60a_0$, The electric fields are respectively $0F_{eff}$, $0.3F_{eff}$, $0.6F_{eff}$ and $0.9F_{eff}$. }
\end{figure}

\begin{figure}
  \includegraphics[width=450pt]{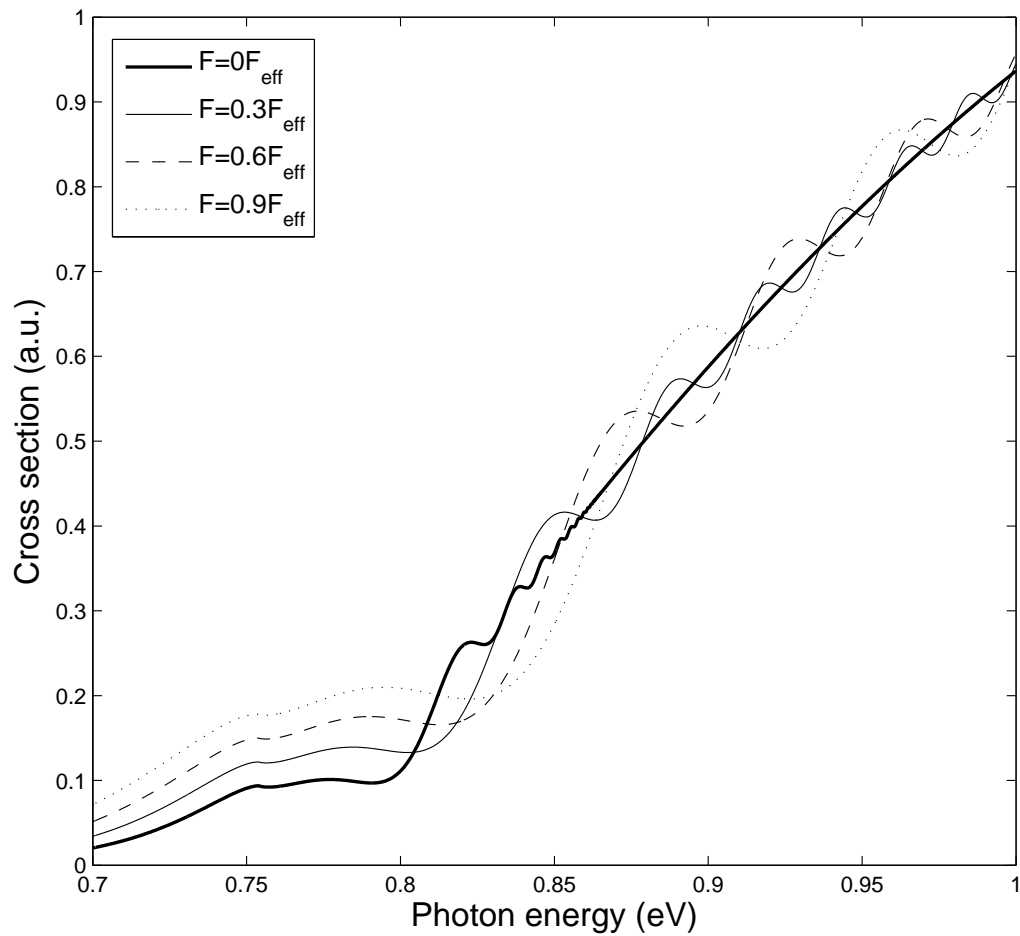}
  \caption{The photodetachment cross sections of H$^-$ in the presence of a metal surface and a static electric field. The distance between the negative ion and the metal surface is $60a_0$. The electric fields are indicated.  }
\end{figure}

\begin{figure}
  \includegraphics[width=450pt]{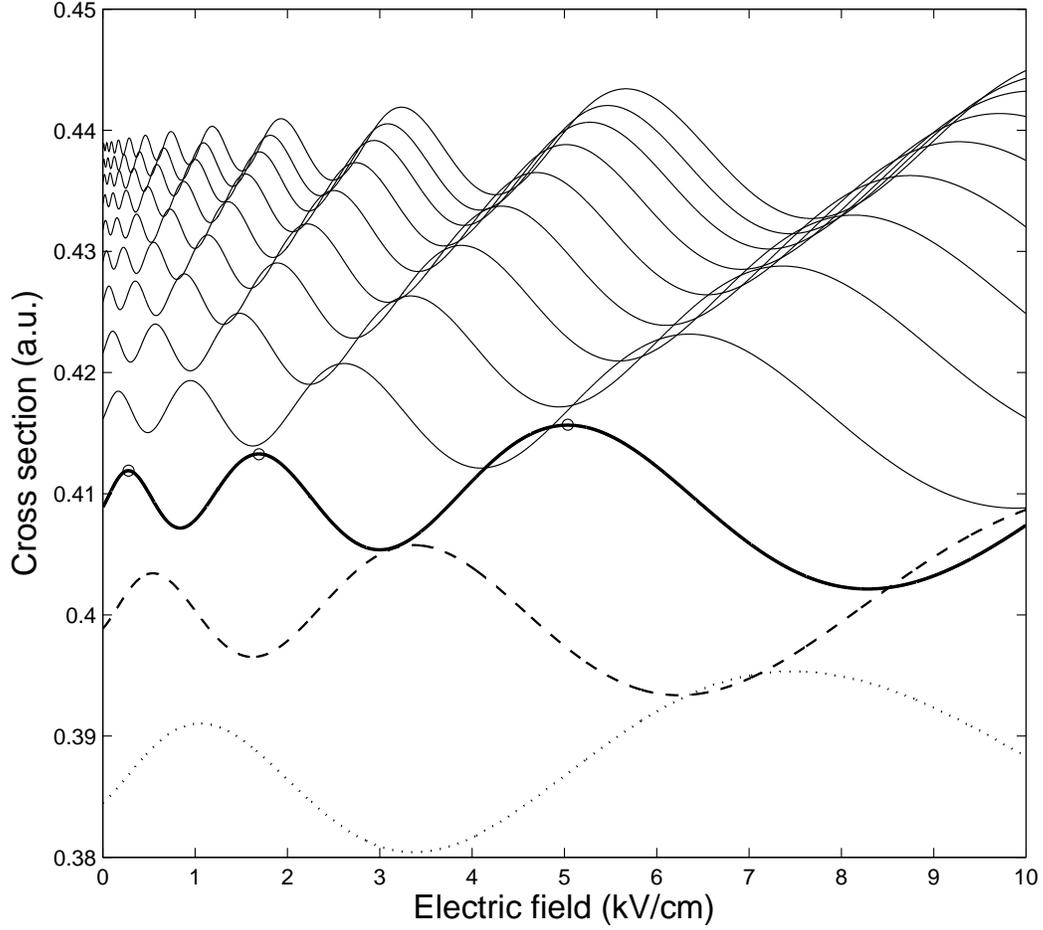}
  \caption{The photodetachment cross sections vs. the electric field. Each curve corresponds to a fixed photon energy. The selected photon energies are larger than $0.85$eV and they correspond to the minima in the cross section when $F$ is set to zero. The photon energies of the first three curves
  are $0.8533$eV, $0.8564$eV and $0.8585$eV which correspond to the dotted line, the dashed line and the heavy solid line. The distance between the negative ion and the metal surface is $60a_0$. }
\end{figure}

\begin{figure}
  \includegraphics[width=450pt]{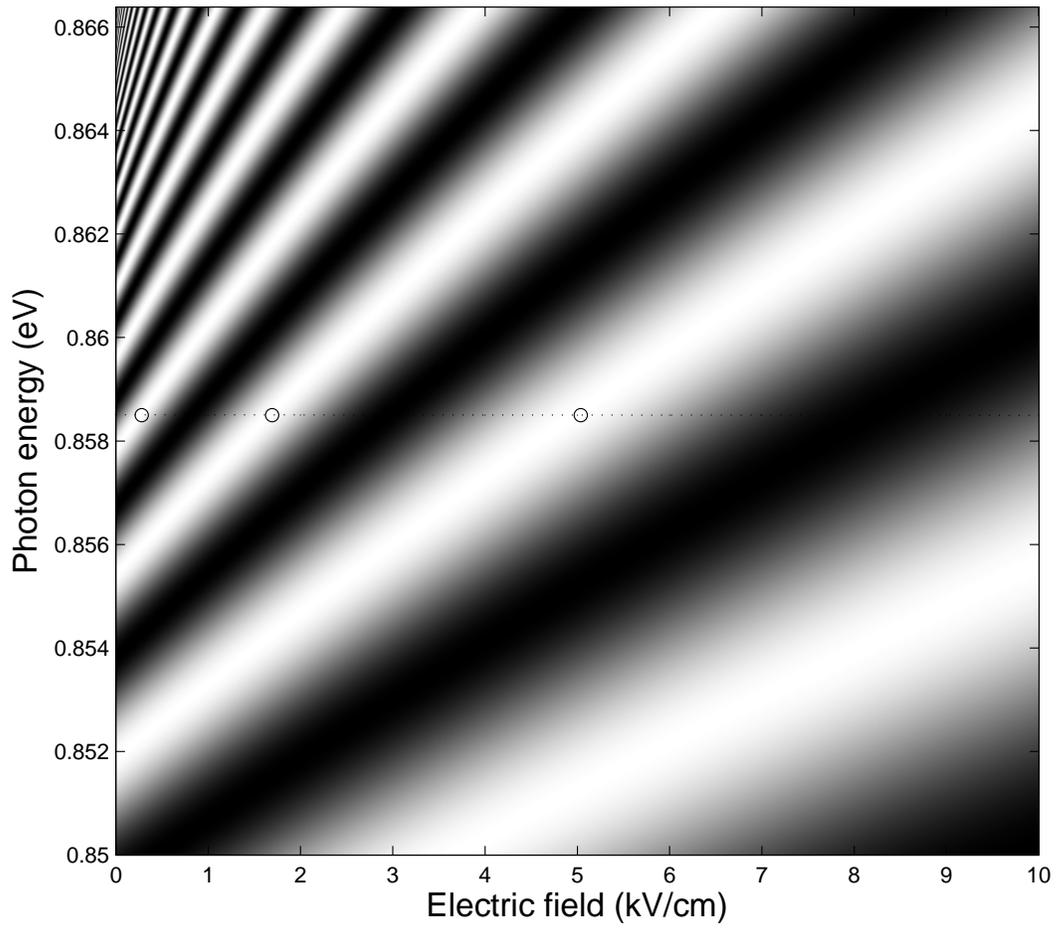}
  \caption{The gray scale function $\cos(S)$ on the photon energy and the electric field plane. The distance between
  the negative ion and the metal surface is $60a_0$. The brightness along the dotted line can be compared with the cross section represented by the heavy solid line in Fig.5. }
\end{figure}

\begin{figure}
  \includegraphics[width=450pt]{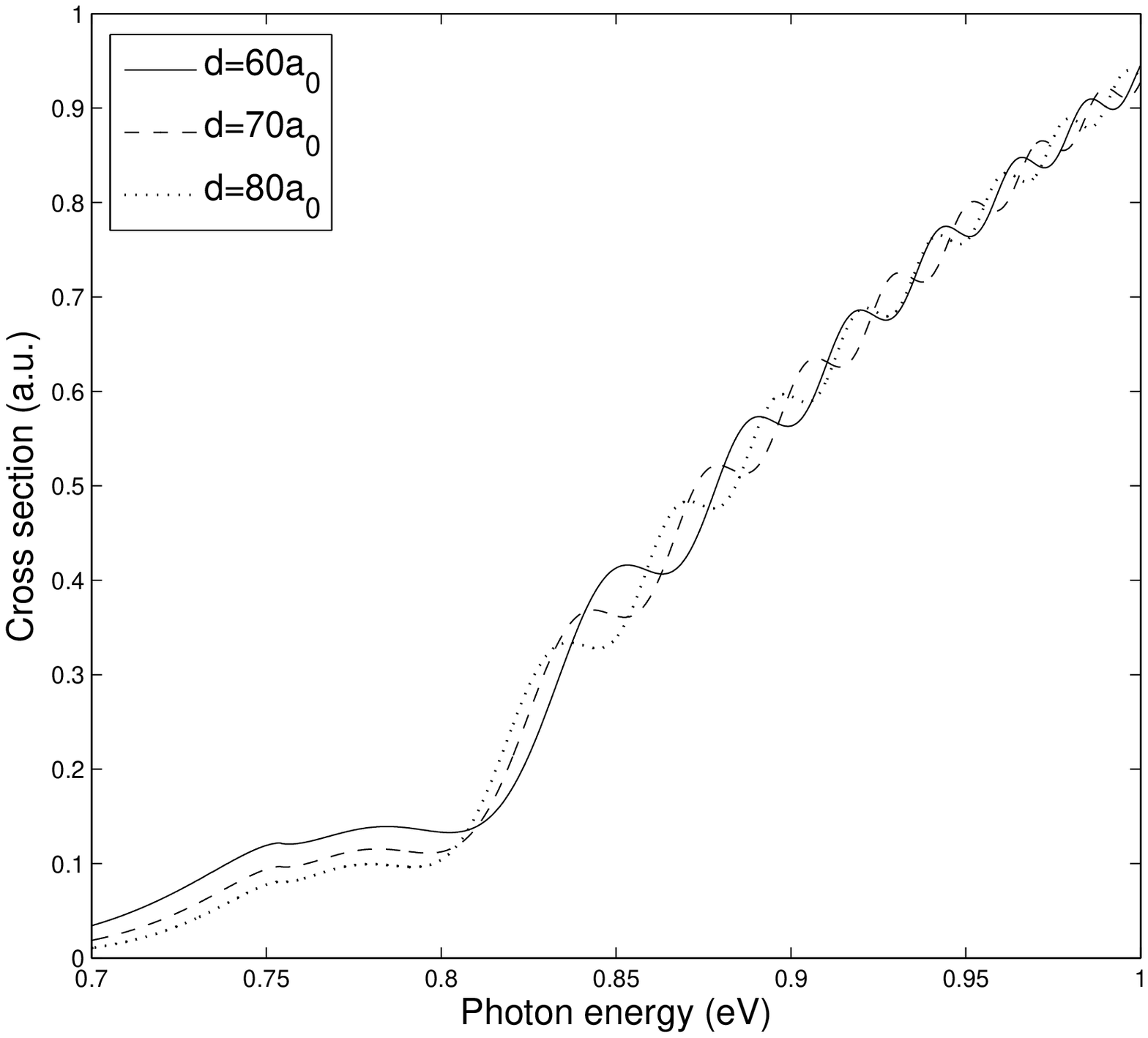}
  \caption{The photodetachment cross sections for three different distances between the negative ion and the metal surface. The static electric field is 107kV/cm which is equal to $0.3F_{eff}$ in Fig.4. The light solid curve in this figure is the same as the light solid curve in Fig.4. }
\end{figure}

\end{center}


\newpage

\begin{references}

\bibitem{Fabrikant1}
I. I. Fabrikant, Sov. Phys.-JETP {\bf 52}, 1045 (1980).

\bibitem{Bryant1}
H. C. Bryant $et$ $al$., Phys. Rev. Lett. {\bf 58}, 2412 (1987).

\bibitem{Stewart}
J. E. Stewart $et$ $al$., Phys. Rev. A {\bf 38}, 5628 (1988).

\bibitem{Rau1}
A. R. P. Rau and H. Wong, Phys. Rev. A {\bf 37}, 632 (1988).

\bibitem{Du1}
M. L. Du and J. B. Delos, Phys. Rev. A {\bf 38}, 5609 (1988); M. L.
Du, Phys. Rev. A {\bf 40}, 1330 (1989).

\bibitem{Gibson1}
 N. D. Gibson, B. J. Davis, and D. J. Larson, Phys. Rev. A {\bf 47}, 1946
 (1993);  N. D. Gibson, B. J. Davis, and D. J. Larson, Phys. Rev. A {\bf 48}, 310 (1993).

\bibitem{Peters}
 A. D. Peters and J. B. Delos, Phys. Rev. A {\bf 47}, 3020 (1993); {\bf 47}, 3036
 (1993);
 A. D. Peters, C. Jaff$\acute{\textrm{e}}$, and J. B. Delos, Phys. Rev. A {\bf 56}, 331 (1997);
 Phys. Rev. Lett. {\bf 73}, 2825 (1994).

\bibitem{Blondel}
C. Blondel, C. Delsart and F. Dulieu, Phys. Rev. Lett. {\bf 77},
3755 (1996).

\bibitem{Rangan}
C. Rangan and A. R. P. Rau, Phys. Rev. A {\bf 61}, 033405 (2000).

\bibitem{Gibson2}
 N. D. Gibson, M. D. Gasda, K. A. Moore, D. A. Zawistowski, and C. W. Walter, Phys. Rev. A {\bf 64}, 061403(R) (2001).

\bibitem{Fabrikant2}
I. I. Fabrikant, M. V. Frolov, N. L. Manakov, and A. F. Starace,
Phys. Rev. A {\bf 64}, 037401 (2001).

\bibitem{Rau2}
 A. R. P. Rau and C. Rangan, Phys. Rev. A {\bf 64}, 037402 (2001).

\bibitem{Kramer}
T. Kramer, C. Bracher and M. Kleber, Euro. phys. Lett. {\bf 56}, 471
(2001)

\bibitem{Du2}
M. L. Du, Phys. Rev. A {\bf 70} 055402 (2004).

\bibitem{Bracher}
C. Bracher and J. B. Delos, Phys. Rev. Lett. {\bf 96}, 100404(2006);
C. Bracher, T. Kramer and J. B. Delos, Phys. Rev. A {\bf 73}, 062114
(2006).

\bibitem{Du3}
 M. L. Du, Eur. Phys. J. D {\bf 38}, 533 (2006).

\bibitem{Ganesan}
K. Ganesan and K. T. Taylor, J. Phys. B: At. Mol. Opt. Phys. {\bf
29} 1293 (1996).

\bibitem{Simonovic1}
N. S. Simonovic, J. Phys. B: At. Mol. Opt. Phys. {\bf 30} L613
(1997).

\bibitem{Salas}
J. P. Salas and N. S. Simonovic, J. Phys. B: At. Mol. Opt. Phys.
{\bf 33} 291 (2000).

\bibitem{Simonovic2}
N. S. Simonovic, Phys. Lett. A {\bf 331} 60 (2004).

\bibitem{Wang}
Dehua Wang, M. L. Du and Shenlu Lin, J. Phys. B: At. Mol. Opt. Phys.
{\bf 39} 3529 (2006).

\bibitem{Yang}
Guangcan Yang \emph{et al}, J. Phys. B: At. Mol. Opt. Phys.
{\bf 39} 1855 (2006).

\bibitem{Zhao1}
Haijun Zhao, Ph.D. thesis, Academia Sinica, 2007.

\bibitem{Zhao2}
H. J. Zhao and M.L.Du, Phys. Rev. A {\bf 79}, 023408 (2009).

\bibitem{Du4}
M. L. Du and J. B. Delos, Phys. Rev. Lett. {\bf 58}, 1731 (1987).

\bibitem{Du5}
 M. L. Du and J. B. Delos, Phys. Rev. A {\bf 38}, 1896 (1988); {\bf 38}, 1913 (1988).
\bibitem{table} We have used the integral formulas 3.141.11 and 3.141.29 in \emph{Table of Integrals,Series, and Products} by I.S.Gradshteyn and I.M.Ryzhik, (Academic Press, New York, 1980).



\end{references}
\end{document}